\documentclass[preprint2]{aastex}
\usepackage{bm,spr-astr-addons}

\slugcomment{Published in Astrop\bfhys. Space Sci. {358:32} No. 2, 1-7 (2015)}

\shorttitle{Astropys. Space Sci. }
\shortauthors{Kawabata}

\begin{document}

\title{A Fast Iterative Method  for  Chandrasekhar's $\bm{H}$-functions\\
    for General Laws of Scattering}


\author{Kiyoshi Kawabata\altaffilmark{}}
\affil{Department of Physics,  Tokyo University of Science,
    Shinjuku-ku, Tokyo 162-8601\\
    E-mail: kawabata@rs.kagu.tus.ac.jp}

\email{kawabata@rs.kagu.tus.ac.jp}

\begin{abstract}
This work shows that notable  acceleration of the speed of calculating Chandrasekhar's $H$-functions for general laws of scattering with an  iterative method can be realized 
 by supplying a starting  approximation produced by the following procedure: (i) in the cases of azimuth-angle independent Fourier components,  values of the isotropic scattering $H$-function given by an accurate yet  simple-to-apply formula, in particular,  the one by \cite{kaw11}, and (ii) for azimuth-angle dependent Fourier components, an already obtained solution of the next lower order term.
\end{abstract}


\section{Introduction}
The intensities of sunlight reflected by a plane-parallel planetary atmosphere
can be expanded in a Fourier series of azimuth-angle difference $\Delta\phi$ of the incident and emergent directions of light. Furthermore, if the atmosphere in question  is semi-infinite in optical depth, each Fourier coefficient of the series can be expressed analytically in terms of the Chandrasekhar's $H$-function \citep{cha60,sob75,hul80}  associated with  a  characteristic function
determined  by the scattering phase function of the atmosphere.

 Although it is  rather straightforward nowadays  
to   carry out  multiple light scattering calculations, it is nevertheless  very 
 useful to have analytical representations of the  emergent intensities of reflected or transmitted sunlight at hand for the purpose of numerical accuracy check of, say, a  newly developed computer program for multiple scattering, or 
for gaining some  physical or mathematical insights into actual  problems involving  light scattering.  It may also be important to keep in mind that theoretical  works in some disciplinary areas of physics such as electron transport in condensed matter require numerical values of  the $H$-function with  relatively high accuracy\citep{jab12}.

Each Fourier component  $H^{(m)}(\varpi_0,\mu)$ of  $H$-function satisfies  the following integral equation:
\begin{eqnarray}
H^{(m)}(\varpi_0,\mu)&=&1+\mu H^{(m)}(\varpi_0,\mu) \nonumber \\
& &\hspace{-1cm} \times\int_0^1\!\!\frac{\psi^{(m)}(\mu')}{\mu+\mu'}H^{(m)}(\varpi_0,\mu')d\mu'  \label{eq-1}
\end{eqnarray}
where $m$ designates the order of the Fourier component, $\psi^{(m)}(\mu)$ is the characteristic function specified by  the  scattering law of our interest, $\varpi_0$ the single scattering albedo, and $\mu$ the cosine of a zenith angle of  the direction of  a ray of light incident on or emerging from the top surface of a semi-infinite atmosphere.
The characteristic function $\psi^{(m)}(\mu)$  depends  also on  $\varpi_0$, but will hereafter be omitted for simplicity.
 
A great deal of efforts have been devoted by various authors to the study on   mathmatical nature of this equation, and  to development of  efficient numerical methods to solve Eq.(\ref{eq-1})\cite[see, e.g.,][]{bos83,kaw91,dav08,kaw11} and the references cited therein). 
For isotropic scattering, for which $\psi^{(0)}(\mu)=\varpi_0/2$, a closed form integral representation of the solution is known, so that we can directly calculate its numerical values.  Furthermore, several accurate approximation formulas are now available\citep{hap93,dav08,kaw11}. 

As for  more general types of anisotropic scattering,  the $H$-functions can be expressed in terms of the characteristic roots of an equation involving $\psi(\mu)$ \citep{cha60},
 but the root-finding process is too time-consuming\cite[see, e.g., ][]{kaw91}. Some sort of  successive numerical iterations are essential  to obtain accurate values for their  $H$-functions. In fact, the iterative method (ii) proposed by \cite{bos83} is highly instrumental in generating numerical solutions 
  correct to 11 significant figures or more. 

In view of the important role played by Chandrasekhar's $H$-functions,  it may be of some significance to further improve the efficiency of  the iterative method of \cite{bos83}. Inspired by the work of \cite{hir94} on the  $H$-function for isotropic scattering,  we would  like to  
investigate, in particular, the effect of   initial approximation on the required number of iterations.

\section{Formalism}

As in \cite{bos83}, we employ an alternative form  of Eq.({\ref{eq-1}) to calculate the  values of  $H$-functions using an  iterative procedure:
\begin{eqnarray}
\frac{1}{H^{(m)}(\varpi_0,\mu)}&=&(1-2\psi^{(m)}_0)^{1/2}+\nonumber \\
& &\hspace{-1.5cm}+\int_0^1\!\!\frac{\mu'\psi^{(m)}(\mu')H^{(m)}(\varpi_0,\mu')}{\mu+\mu'}d\mu',  \label{eq-2}
\end{eqnarray} 
where we have put 
\begin{equation}
\psi^{(m)}_0=\int_0^1\!\!\!\psi^{(m)}(\mu')d\mu'.    \label{eq-3}
\end{equation}

For numerical calculations, we replace the integral with respect to $\mu'$ over the interval $\mu'\in[0, 1]$ with a quadrature of degree $N$. In the present work, the $N_{\rm G}$-point Gauss-Legendre  quadrature is chosen  for this purpose, so that Eq.(\ref{eq-2}) takes the following form:
\begin{eqnarray}
\frac{1}{H^{(m)}(\varpi_0, \mu_k)} &=(1-2\psi^{(m)}_0)^{1/2}+& \nonumber \\ 
& &\hspace{-4.5cm}\displaystyle{+\sum_{j=1}^{N_{\rm G}}\frac{\mu_j\psi^{(m)}(\mu_j)H^{(m)}(\varpi_0,\mu_j)}{\mu_k+\mu_j}w_j  } \label{eq-4}\\
& (k=1,\dots, N_{\rm G}) \nonumber
\end{eqnarray}
with $\mu_j$ and $w_j$ being the $j$-th quadrature point and  the  corresponding weight, and $\mu_k$ the $k$-th quadrature point at which the numerical solution is sought for.

In order to study the effect of anisotropy of scattering on each method employed in this work,  a four-term phase function of the form
\begin{equation}
P(\cos\Theta)=\varpi_0\left\{\displaystyle{1+\sum_{j=1}^{J}x_j P_j(\cos\Theta)}\right\}  \quad (J\le 3)\label{eq-5}
\end{equation}
is  employed,  where $P_j(\cos\Theta)$ is the Legendre polynomial function of the $j$-th degree, $x_j$'s  are the expansion coefficients, and $\Theta$ is the scattering angle \citep{hul80}. $J$ is the maximum degree of the Legendre functions to be taken into account, and coincides with the highest degree $M$ of the Fourier terms required to represent the azimuth-angle dependence of emergent intensities of reflected light, which is in turn the  highest degree of the Fourier components $H^{(m)}(\varpi_0,\mu)$: for isotropic scattering, where  $J=0$, only the  azimuth-angle independent Fourier component $H^{(0)}(\varpi,\mu)$ needs to be considered,  which we shall refer to as ${}^{\rm iso}H(\varpi_0,\mu)$. This four-term phase function is convenient for  the fact that it covers,  as special cases, 
\begin{enumerate}
\item[(i)] isotropic scattering phase function given by $x_j=0 \ (j=1,\dots, 3)$, 
\item[(ii)] Rayleigh scattering phase function with $x_1=0$, $x_2=1/2$, and $x_3=0$, 
\item[(iii)] linearly anisotropic scattering phase functions with $x_1\ne 0$, and $x_2=x_3=0$,
\item[(iv)] three-term phase functions with $x_3=0$, 
\item[(v)] four-term phase functions with $x_3\ne 0$.
\end{enumerate} 

The $m$-th order Fourier components $\psi^{(m)}(\mu)\ (m=0, 1, 2, 3)$ corresponding to  the four-term  phase function Eq.(\ref{eq-5})  are  as follows: 
\begin{mathletters}
\begin{eqnarray}
m=0\!\!\!\!&:&\!\!\!\!\psi^{(0)}(\mu)=\frac{1}{2}\varpi_0\left\{1+\frac{1}{4}x_2+\left[h_0x_1-\frac{3}{4}x_2  \right.\right. \nonumber \\ 
& &\left.-\frac{1}{4}h_0h_1x_2 +h_0x_3+\frac{1}{4}h_2x_3\right]\mu^2 \nonumber\\ 
& &+\left[\frac{3}{4}h_0h_1x_2-\frac{5}{3}h_0x_3-\frac{5}{12}h_2x_3 \right.\nonumber \\
& &\left.\left. -\frac{1}{4}h_0h_1h_2x_3\right]\mu^4+\frac{5}{12}h_0h_1h_2x_3\mu^6\right\}\\
m=1\!\!\!\!&:&\!\!\!\!\psi^{(1)}(\mu)=\frac{1}{2}\varpi_0(1-\mu^2)\left\{\frac{1}{2}x_1+\frac{3}{16}x_3 \right. \nonumber \\
& &+\left[\frac{1}{2}h_1x_2-\frac{1}{16}(h_1h_2+15)x_3\right]\mu^2\nonumber \\
& &\left.+\frac{5}{16}h_1h_2x_3\mu^4\right\}\\
m=2\!\!\!\!&:&\!\!\!\!\psi^{(2)}(\mu)=\frac{3}{16}\varpi_0(1-\mu^2)^2 \nonumber\\
& &\times(x_2+x_3h_2\mu^2) \\
m=3\!\!\!\!&:&\!\!\!\!\psi^{(3)}(\mu)=\frac{5}{32}\varpi_0x_3(1-\mu^2)^3
\end{eqnarray}
\label{eq-6}
\end{mathletters}   
\hspace*{-0.3cm}where $h_k\ \ (k=0, 1, 2,3)$ is defined as
\begin{equation}
h_k=2k+1-\varpi_0x_k,  \label{eq-7}
\end{equation} 
following the Display 6.2 of \cite{hul80}.

As was initially recognized  by \cite{cha60} and  later demonstrated mathematically  by Bosma \& de Rooij (1983),  
Eq.(\ref{eq-2}) and hence Eq.(\ref{eq-4}) is more suited for  iterative methods, and in fact, 
tabulation of the values of the $H$-functions for  certain laws  of  anisotropic scattering have been made by  e.g.,  \cite{cha60}, \cite{kol72}, and Bosma \& de Rooij (1983)  solving this alternative form of the basic equation. 
A recourse is correspondingly made to Eq.(\ref{eq-4}) also in the present  work. 

For a given value of $\varpi_0$, a convergence to the  solution of Eq.(\ref{eq-4}) is assumed to have been realized  after the $n$-th iteration, if the following condition is satisfied:
\begin{eqnarray}
\max& \left|H^{(m)}(\varpi_0,\mu_j)_n-H^{(m)}(\varpi_0,\mu_j)_{n-1}\right|\le \varepsilon, \nonumber \\
& (j=1, 2, 3, \dots, N_{\rm G})  \label{eq-8}
\end{eqnarray}
where we adopt absolute-error tolerance  $\varepsilon=10^{-12}$ as in \cite{bos83} to ensure  11 significant figure accuracy.
If, on the other hand,  this condition is not fulfilled, we have two alternatives for creating a  next approximation $\overline{H}(\varpi_0,\mu)_n$ 
for $H^{(m)}(\varpi_0,\mu)$ on the right  side of Eq.(\ref{eq-4})    before proceeding to the $(n+1)$-th iteration: 
\begin{equation}
(1)\  \overline{H}(\varpi_0,\mu)_n=H^{(m)}(\varpi_0,\mu_j)_n/H^{(m)}(\varpi_0,0)_n.  \label{eq-9}
\end{equation}
The values for $H^{(m)}(\varpi_0,0)_n$ required above are to be calculated at the end of the $n$-th iteration using Eq.(\ref{eq-4}) as 
\begin{eqnarray}
\frac{1}{H^{(m)}(\varpi_0, 0)_n} &=(1-2\psi^{(m)}_0)^{1/2}+& \nonumber \\ 
& &\hspace{-4.5cm}\displaystyle{+\sum_{j=1}^{N_{\rm G}}\psi^{(m)}(\mu_j)H^{(m)}(\varpi_0,\mu_j)_n w_j  } \label{eq-10}\\
&  (k=1,2,3, \dots, N_{\rm G})& \nonumber
\end{eqnarray}
or
\begin{eqnarray}
(2)\ {\overline{H}}^{(m)}(\varpi_0,\mu)_n&=&\lambda_0 H^{(m)}(\varpi_0,\mu_j)_n\hspace{1.3cm}\nonumber \\
& &\hspace{-1.5cm}+(1-\lambda_0)H^{(m)}(\varpi_0, \mu_j)_{(n-1)}     \label{eq-11}
\end{eqnarray}
\cite{bos83} suggested, in their method (i),  the use of 
\begin{equation}
\lambda_0=\frac{1}{2}\left[1+(1-2\psi^{(m)}_0)^{1/2}\right], \label{eq-12}
\end{equation}
Eq.(\ref{eq-12}) yields $\lambda_0=0.5$ only for the conservative case where $\varpi_0=1$. \cite{hir94}, on the other hand, employed  $\lambda_0=0.5$ exclusively to solve for the isotropic scattering $H$-function for arbitrary value of $\varpi_0$.

\section{Numerical Calculations}
Table~\ref{htbl-1} shows the data set used for the present work, compiling  37 cases of phase functions  that can be represented by the the four-term phase function given by Eq.(\ref{eq-5}). The last column of the table indicates an abbreviated name of  the scattering law characterizing  each entry: the row No.1  designated by 'ISO' is for  isotropic scattering,  the rows No.2 through No.7 designated by 'LIN' are  linearly anisotropic scattering phase functions, each of which yields  an azimuth-angle independent  component $\psi^{(0)}(\mu)$ and a first order azimuth-angle dependent  Fourier component $\psi^{(1)}(\mu)$. The row No.8 designated by 'RAY' is for Rayleigh scattering, a special case of a  three-term phase function,  and gives rise to 
 three Fourier component characteristic functions $\psi^{(m)}(\mu)\ \ (m=0, 1, 2)$, the rows No. 9 through  No.23 are for more general cases of the three-term phase function ($J=2$),  each having  three Fourier components as in No.8, and  the rows No.24 through No.37 are for four-term phase functions ($J=3$),   each bearing  four Fourier component  characteristic functions   $\psi^{(m)}(\mu) \ \ (m=0, 1, 2, 3)$.  As the result, altogether 117 cases of characteristic functions arise, for each of which  Eq.(\ref{eq-4}) is sovled iteratively for $H^{(m)}(\varpi_0,\mu)$.  Asterisked entries  indicate that their  phase functions show  negative values at some  scattering angles due to insufficient degree of approximation obtained with the first four Legendre polynomials.

Following \cite{bos83}, we also employ $N=N_{\rm G}=128$ for the Gauss-Legendre quadrature, and $\varepsilon=10^{-12}$ (see Eq.(\ref{eq-8})). For each of the 117 cases of  the Fourier components $H^{(m)}(\varpi_0,\mu)$
given by the 37 phase functions enumerated in Table~\ref{htbl-1}, calculations of $H^{(m)}(\varpi_0,\mu)$\ $(m\le 3)$ are carried out solving  Eq.(\ref{eq-4}) iteratively at mesh points $(\varpi_0,\mu)$ specified  by  14 values of   $\varpi_0$, viz., $10^{-3}$,  $0.05$, $0.1$ through $0.9$ ( with a step of $0.1$), $0.99$, $0.999$, and $1$, and 128 quadrature points $\mu_k$ $(k=1, 2, 3, \dots,128)$. Once a set of  numerical solution  $H^{(m)}(\varpi_0,\mu)$ has been found at 128 quadrature points for a given value of $\varpi_0$, 
we can freely calculate the value of $H^{(m)}(\varpi_0,\mu)$ at more regularly spaced non-quadrature points $\mu$  by replacing $\mu_k$ with $\mu$ in  Eq.(\ref{eq-4}).  For convenience of comparison with the results given in other literature, we employ  21 values for $\mu$, viz., 0 through 1 with a step of 0.05, for tabulation.

In order to access the efficiency of   each scheme we examine,  we have proceeded in the  following four steps successively:\\

\noindent{\bf Method A}: \\
To produce  a control case, the numerical solutions  for  $H^{(m)}$ $(\varpi_0,\mu)$ are  calculated for the 117 cases of characteristic functions   by applying  the method (ii) of \cite{bos83}:
the starting approximation is simply $\overline{H}^{(m)}(\varpi_0,\mu)_0=1$   irrespective of  the characteristic function involved.  Furthermore, the successive iteration ($n\ge 1$) is carried out with  Eq.(\ref{eq-9}).

To verify  our computer code,  our numerical values for the cases No.1, No.2, and No.8 given in Table~\ref{htbl-1} were compared with those     
 in Tables 1a through 1f of \cite{bos83}, to find 
 that both results are in perfect  agreement to each other to the last decimal figures. For No.1, the number of iterations $N_{\rm it}$ required to get the numerical solution for $H^{(0)}(\varpi_0,\mu)$  were  also in fair  agreement with the corresponding result given in their Table 3.  \\

\noindent{\bf Method B}:\\
   Inspired by the work of \cite{hir94} for  isotropic scattering, we conceived of a possibility that 
supplying a significantly more accurate  starting approximation, we  might be able to further  increase  the efficiency of  the iterative method  even for anisotropic scattering phase functions. One of the simplest procedures to realize this would  be to  introduce a more  improved approximation formula 
than that of   \cite{hap93} adopted in \cite{hir94}
 to generate the values of    ${}^{\rm iso}H$ without requiring any laborious numerical calculations..
 
From this stand point,   two excellent approximation formulas are now available: (a) an analytic formula derived by \cite{dav08}, and (b) a rational approximation formula obtained by \cite{kaw11} based on a least-squares fitting to an accurate set of tabulated values of ${}^{\rm iso}H(\varpi_0,\mu)$. 
The  maximum relative errors  of  (a) and (b) are  $0.07$ \% and $2.1\times 10^{-4}$ \%, respectively, whereas  that of Hapke's formula is about  $0.8$ \%.   Let us select  the formula (a) here and leave the formula (b) for
later comparison. 
   
For the $(n+1)$-th iteration ($n\ge 1$), a new approximation $\overline{H}^{(m)}(\varpi_0,\mu)_n$ is given  by Eq.(\ref{eq-11}) together with  the acceleration parameter $\lambda_0$  calculated by Eq.(\ref{eq-12}) following the  method (i) of \cite{bos83}.  \\

\noindent{\bf Method C}:\\
  As in the Method B, the analytic formula of Davidovic' et al. (1980) 
   is employed to create  starting approximations to initiate the iterative solutions for all of the 117 cases originating  from  Table~\ref{htbl-1}.   However,  Eq.(\ref{eq-9}) rather than  Eq.(\ref{eq-11}) is applied to obtain a new approximation $\overline{H}^{(m)}(\varpi_0,\mu)_n
  $ for carrying out the $(n+1)$-th iteration ($n\ge 1$). 
  
As our experiments progressed, it became increasingly clear that the use of an accurate formula for ${}^{\rm iso}H(\varpi_0,\mu)$ as a starting approximation generator can help  improve the efficiency of calculating the values of the azimuth-angle independent components $H^{(0)}(\varpi_0,\mu)$ even for general laws of scattering.

 For $m\ge 1$, on the other hand, such  technique  did  not  seem to be working as effectively as in the $m=0$ cases,
 and something  else   was desired   to  speed up the calculations of $H^{(m)}(\varpi_0,\mu)$.\\

\noindent{\bf Method D}:\\
   The rational approximation formula of Kawabata \& Limaye (2011)  
    \citep[see also ][for erratum]{kaw13} for ${}^{\rm iso}H(\varpi_0,\mu)$ rather than that of Davidovic' et al. (2008) 
     is employed to produce the starting approximation to solve  Eq.(\ref{eq-4}) in the $m=0$ cases.
The formula we are going to use is of the following form:
\begin{equation}
\displaystyle{H(\varpi_0,\mu)=H_{\rm app}(1,\mu)/[1+\sum_{k=0}^8C_k(\varpi_0)x^k]      } \label{eq-13}
\end{equation}
where  $x=\mu^{1/4}$, while $H_{\rm app}(1,\mu)$ and $C(\varpi_0)$ are  eighth order polynomials of $x$ and $\sqrt{1-\varpi_0}$, respectively.  As has already been mentioned, this formula is based on extensive  numerical data of ${}^{\rm iso}H(\varpi_0,\mu)$ produced with 11 digit accuracy by \cite{kaw11} using a closed form solution.
\footnote{\cite{jab12} lately cast doubt on the numerical accuracy of the method developed  by \cite{kaw11}: he erroneously states that the Kawabata-Limaye method yielded  $H(1,1)=2.9077901976$, while the  more correct value is
   $2.9078105291$. On the contrary,  their method is capable of generating the values of ${}^{\rm iso}H(\varpi_0,\mu)$  accurate at least to the 10th decimal figures for any combination of $\varpi_0$ and $\mu$ values as their Table 1  clearly shows.  The value $2.9078105291$ was mentioned in \cite{kaw11} simply  to warn the readers that such less accurate figures would  result unless their method or something alike is employed.}

For the $m (\ge 1)$-th order Fourier components,  however,  we substitute  $H^{(m-1)}(\varpi_0,\mu)$, the solution for the $(m-1)$-th Fourier component, as the initial approximation for  iterations. 
This procedure, though simple as it is,  was found fairly promising during some preliminary experiments.
 The approximation  $\overline{H}(\varpi_0,\mu)_n$ used for the $(n+1)$-th iteration is calculated by  Eq.(\ref{eq-9}).

\section{Results}
The left half of Table \ref{htbl-2} compares  in the case of isotropic scattering the numbers of iterations $N_{\rm it}$ required for four methods A, B, C, and D to achieve  a convergence  within $\varepsilon=10^{-12}$ to the solution of Eq.(\ref{eq-4}) for 14 values of $\varpi_0$.
Also shown in each row of  the right half of the table are the values for $N_{\rm it}-\min N_{\rm it}$ for the same four methods, where $\min N_{\rm it}$ designates the minimum of the four $N_{\rm it}$ values given in the same  row on the left. 
The bottom row  shows  sum of these quantities given in each column.
We notice  that  Eq.(\ref{eq-13}) derived by  \cite{kaw11} enables us to cut down  the number of iterations $N_{\rm it}$ by a factor of more than two in the  cases of isotropic scattering.

Of course, the true capability of each method must be 
assessed from the stand point of  aniso\-tropic scattering calculations. 
  The topmost row of Table~\ref{htbl-3}  shows the values of  $N_{\rm it}-\min N_{\rm it}$ required to calculate 116 sets of solutions for $H^{(m)}(\varpi_0,\mu)\ (m\ge 0)$ arising from  the phase functions No.2 through No.37 of Table~\ref{htbl-1}. The second and   third rows are a breakdown 
of the statistics given in the first row  into two groups:  the figures  shown in the second row are for 37 azimuth-angle independent components $H^{(0)}(\varpi_0,\mu)$, and those  shown in the third row are 
 for 79 cases  of azimuth-angle dependent Fourier components $H^{(m)}(\varpi_0,\mu)$\ ($m\ge 1$),  respectively.
The parenthesized figures  in each row indicate 
 fractional contributions  from  the four methods tested in this work to the total value of $N_{\rm it}-\min N_{\rm it}$ given in its  last column. The performance of the method D is definitely outstanding as far as these statistics go.
 
Table~\ref{htbl-4}  presents, as an example, the values obtained with the method D for  $H^{(m)}(1,\mu) \ (m=0, 1, 2,3)$ corresponding to the conservative scattering case due to the phase function No.34 of Table~\ref{htbl-1}  as functions of 21 $\mu$ values.  The number of iterations $N_{\rm it}$ necessary  to get these solutions are given in the bottom row.  The  last column of the table shows the resulting values of the azimuth-angle independent component of  reflection function for $\mu=\mu_0$ or $R^{(0)}(1,\mu,\mu)$, where the reflection function is defined as
\begin{equation}
 R^{(0)}(\mu,\mu_0)\equiv I^{(0)}(\mu,\mu_0)/(F_0\mu_0), \label{eq-14}
\end{equation} 
with $I^{(0)}(\mu,\mu_0)$ being the azimuth-angle averaged intensity of reflected light emerging from a semi-infinite atmosphere in the direction specified by a zenith angle $\theta=\cos^{-1}\mu$ originating from the sunlight incident from a direction with a zenith angle $\theta_0=\cos^{-1}\mu_0$.
 The quantity $F_0$, if multiplied by $\pi$,  designates the incident radiative flux per unit area perpendicular to the incident direction. 
 
We have made a  comparison of our results} for $H^{(0)}(1,\mu)$ with those given in Table 3 of  Kolesov \& Smoktii (1972) 
to find agreement  to the third decimal places except that  differences by one unit  in the third decimal places  are observed  at several scattered locations probably due to a round-off effect: incidentally, no such  difference is present in the corresponding values of Table~\ref{htbl-4}. 
A comparison has also been made between our values of  $R^{(0)}(1,\mu,\mu)$ and those shown in Table 4 of \cite{kol72}: both calculations largely  agree  to the third decimal places except at some entries where discrepancies mostly by one unit in the third decimal figures are found, the largest of which occurs for the reflection function for isotropic scattering at $\mu=0.5$, where we have 1.0128195942 in contrast to 1.023 obtained by \cite{kol72}. It may be  worth noting that our value is in agreement with 1.01282 shown in Table 12 of \cite{hul80}.
 
\section{Conclusion}
On the basis of the numerical experiments carried out  in this work in accordance with  the four methods A, B, C, and D,  we have reached the following conclusions:
\begin{enumerate}
\item[(i)] The iterative method to solve Eq.(\ref{eq-2}) for general laws of scattering works quite efficiently, if the normalization procedure  of \cite{bos83}, Eq.(\ref{eq-9}), is applied to create an approximate solution for  $H$-function prior to each successive iteration.  The efficiency of the iterative solution can further  be improved by supplying adequate starting  numerical approximations to initiate the iteration.   
\item[(ii)]For azimuth-angle independent components, the number of iterations required to obtain solutions
for  $H^{(0)}(\varpi_0,\mu)$ can be significantly reduced  by using,  as a starting approximation, the  ${}^{\rm iso}H(\varpi_0,\mu)$ generated with  the approximation formula obtained by  \cite{kaw11} whose maximum relative error is $2.1\times 10^{-4}$ \%.
\item[(iii)] In the cases of  higher order Fourier  components $H^{(m)}(\varpi_0,\mu)\ (m\ge 1)$,  the substitution  of  the  $(m-1)$-th order solution  $H^{(m-1)}(\varpi_0,\mu)$ as the initial approximation greatly reduce  the number of iterations necessary. \footnote{A Fortran77 source program to calculate $H$-functions applying the method D is available from the author on request.}
\end{enumerate}


\acknowledgments
\noindent{\bf Acknowledgments}~The author is grateful to the anonymous referee for his or her 
constructive comments.\par
This has been published in Astrophysics and Space Science (2015) {\bf 358:32}, 1-7, DOI 10.1007/\\
s10509-015-2434-0. The final publication is available at link.springer.com.

\clearpage

\begin{table}
\begin{center}
\caption{Phase functions employed for numerical experiments\label{htbl-1}}
\begin{tabular}{crrrrr}
\tableline\tableline
No. &$J$ &  $x_1$ & $x_2$ & $x_3$& $\mbox{ref}^\dagger$ \\
\tableline
1)&0  &  0&  0 &  0 & ISO  \\
2)&1 &   1  &  0 &    0 & LIN \\
3)&1  &  0.9 &  0 &   0 & LIN\\
4)&1  &  0.5&  0&   0 & LIN\\
5)&1 &  -0.5 &  0 &   0 & LIN\\
6)&1 &  -0.9 &  0 &   0 & LIN\\
7)&1 &  -1 &  0 &  0 &LIN\\
8)&2 &   0 &  0.5 &  0 & RAY\\   
9)&2 &   1 &  1 &  0& SOB  \\
10)&2 &   1.5& 1&   0&  SOB \\
11)&2  &  1.076& 0.795& 0&KS3\\
12)&2 &   0.240& 0.498 & 0&KS3\\
13)&2  &  0.092& 0.497& 0&KS3\\
14)*&2 &   2.670& 2.470& 0 &KS3\\
15)&2 &   1.269& 0.909& 0&KS3\\
16)&2  &  0.566& 0.566& 0&KS3\\
17)*&2  &  2.879&2.740&0&KS3\\
18)&2 &   1.198& 0.869& 0&KS3\\
19)&2 &   0.540&0.568& 0&KS3\\
20)*&2 &   2.560& 2.285 &0&KS3\\
21)*&2 &   1.789&1.265& 0&KS3\\
22)*&2 &   2.698& 2.459& 0&KS3\\
23)*&2 &   1.759&1.283& 0&KS3\\
24)&3 &  1.006&0.795& 0.215&KS4\\
25)&3 &  0.208&0.498& 0.098&KS4\\
26)&3 &  0.083& 0.497& 0.028&KS4\\
27)*&3 &  1.972& 2.470& 1.635&KS4\\
28)&3 &  1.180& 0.909&0.269&KS4\\
29)&3 &  0.529& 0.566& 0.113&KS4\\
30)*&3 &  2.079& 2.740& 1.875&KS4\\
31)&3 &  1.110&0.869& 0.266&KS4\\
32)&3 &  0.510&0.568& 0.092&KS4\\
33)*&3&1.948& 2.285& 1.432& KS4\\
34)&3 &  1.615& 1.266& 0.432&KS4\\
35)*&3 &  2.028&2.450& 1.569&KS4\\
36)&3 &  1.560&1.283& 0.494&KS4\\
37)&3 &  0 & 1&1&HUL\\
\tableline
\end{tabular}
\tablenotetext{\dagger}{ISO: isotropic scattering; LIN: linearly anisotropic scattering; \\
RAY: Rayleigh scattering; SOB: \cite{sob75}, Table 7.1; \\
KS3, KS4: \cite{kol72}, Table 1;\\
HUL: \cite{hul80}, Table 29}.
\tablenotetext{*}{Phase functions exhibit negative values \\ depending on the values of scattering angle.}
\end{center}
\end{table}

\clearpage

\begin{table}
\begin{center}
\caption{Number of iterations $N_{\rm it}$ required  to calculate $H$-function for  isotropic scattering\label{htbl-2}}
\begin{tabular}{crrrr|rrrr}
\tableline\tableline
&\multicolumn{4}{c|}{$N_{\rm it}$}&\multicolumn{4}{c}{$N_{\rm it}-\min N_{\rm it}$} \\
\tableline
$\varpi_0$ & A & B & C & D& A & B& C& D \\
\tableline
\tableline     
  1.000 & 12 & 31 & 10 &  7 &     5  & 24   & 3  &  0  \\
  0.999 & 12 & 29 &  9 &  6 &     6   &23   & 3   & 0  \\
  0.990 & 12 & 27 & 10 &  6 &     6   &21   & 4  &  0  \\
  0.900 & 13 & 20 & 10 &  6 &     7   &14   & 4  &  0  \\
  0.800 & 14 & 17 & 10 &  6 &     8   &11   & 4  &  0  \\
  0.700 & 15 & 15 & 11 &  5 &    10   &10   & 6 &   0  \\
  0.600 & 14 & 13 & 11 &  6 &     8    &7   & 5  &  0  \\
  0.500 & 13 & 11 & 10 &  5 &     8   & 6   & 5  &  0  \\
  0.400 & 12 & 10 &  9 &  5 &     7   & 5   & 4  &  0  \\
  0.300 & 11 &  9 &  8 &  4 &     7   & 5   & 4  &  0  \\
  0.200 & 10  &  7 &  7 &  4 &     6  &  3   & 3  &  0  \\
  0.100 &  8 &  6 &  6 &  4 &     4   & 2   & 2  &  0  \\
  0.050 &  7 &  5 &  5 &  3 &     4  &  2   & 2  &  0  \\
  0.001 &  4 &  2 &  2 &  2 &     2   & 0   & 0   & 0  \\
\tableline     
  Total & 157    & 202    & 118   & 69    &    88 & 133  & 49  &  0     \\

\tableline
\end{tabular}
\end{center}
\end{table}

\clearpage
\begin{table}
\begin{center}
\caption{$N_{\rm it}-\min N_{\rm it}$ for the 116 cases involving anisotropic phase functions\label{htbl-3}}
\begin{tabular}{cr|rrrrr}
\tableline\tableline
Cases&\#& A & B & C& D  & Total\\
\tableline
$m\ge 0$&116     &981& 5125&  908 &  80 & 7094\\
 &    & (.138)&( .722)& (.128)& (.011)&\\
\tableline 
$m=0$&37& 349 &4180 &  19 &   0 & 4548\\
 &(.319) & (.077)& (.919& (.004)& (.0) &  \\
\tableline
 $m\ge 1$&79&632&  945 & 889 &  80  & 2546\\
&(.681)&(.248)& (.371)&( .349)& (.031) & \\
\tableline 
\end{tabular}
\end{center}
\end{table}

\clearpage
\begin{table}
\begin{center}
\caption{Values of $H$-functions  and azimuth-angle averaged reflection function for conservative scattering with the phase function No. 34 of Table 1\tablenotemark{a} using Method D \label{htbl-4}}
\begin{tabular}{crrrrr}
\tableline\tableline
 $\mu$       &      $H^{(0)}(1,\mu)$ &     $H^{(1)}(1,\mu)$ &   $H^{(2)}(1,\mu)$ &    $H^{(3)}(1,\mu)$ &     $R^{(0)}(\mu,\mu)$\\
\tableline 
 0.00     &1.0000000000 &1.0000000000 &1.0000000000 &1.0000000000 & $\infty$\phantom{11111}    \\
 0.05     &1.1659440619 &1.0771633075 &1.0332050599 &1.0076297119 &4.2285847359 \\
 0.10     &1.2989965575 &1.1265567212 &1.0516671536 &1.0113354601 &2.4817486757 \\
 0.15     &1.4229520561 &1.1661176772 &1.0652788635 &1.0138828020 &1.8799139139 \\
 0.20     &1.5420072951 &1.1995291407 &1.0760596942 &1.0158004425 &1.5711341592 \\
 0.25     &1.6579405618 &1.2285300089 &1.0849344306 &1.0173173607 &1.3827844072 \\
 0.30     &1.7717010913 &1.2541429670 &1.0924264204 &1.0185568495 &1.2569336797 \\
 0.35     &1.8838624879 &1.2770429808 &1.0988669831 &1.0195935779 &1.1687277855 \\
 0.40     &1.9947999590 &1.2977085807 &1.1044812796 &1.0204763205 &1.1057803494 \\
 0.45     &2.1047729686 &1.3164959702 &1.1094300709 &1.0212386882 &1.0612366041 \\
 0.50     &2.2139685305 &1.3336798109 &1.1138324177 &1.0219047912 &1.0309698193 \\
 0.55     &2.3225258489 &1.3494776133 &1.1177790365 &1.0224924770 &1.0122907758 \\
 0.60     &2.4305512527 &1.3640652645 &1.1213406392 &1.0230152976 &1.0032916519 \\
 0.65     &2.5381277033 &1.3775874048 &1.1245733862 &1.0234837620 &1.0024844377 \\
 0.70     &2.6453210934 &1.3901646382 &1.1275225888 &1.0239061654 &1.0085874221 \\
 0.75     &2.7521845597 &1.4018987024 &1.1302252991 &1.0242891558 &1.0203910630 \\
 0.80     &2.8587615184 &1.4128762757 &1.1327121707 &1.0246381324 &1.0366687265 \\
 0.85     &2.9650878522 &1.4231718428 &1.1350088237 &1.0249575309 &1.0561139435 \\
 0.90     &3.0711935192 &1.4328498923 &1.1371368652 &1.0252510332 &1.0772939407 \\
 0.95     &3.1771037571 &1.4419666308 &1.1391146657 &1.0255217236 &1.0986134940 \\
 1.00     &3.2828399994 &1.4505713372 &1.1409579575 &1.0257722074 &1.1182855176 \\
 \tableline
\#iters.\tablenotemark{b}          &          12\phantom{1234} &          14\phantom{1234} &          11\phantom{1234} &          7\phantom{1234}  &                 \\
\tableline
\end{tabular}
\tablenotetext{a}{A four-term phase function with $x_1=1.615$, $x_2=1.266$, and $x_3=0.432$}
\tablenotetext{b}{The numerical figures to the right  indicate the number of iterations needed to get the values of \\ 
$H^{(m)}(1,\mu)\ (m=0, 1, 2, 3)$ with $\varepsilon=10^{-12}$.} 
\end{center}

\end{table}

\end{document}